\documentclass{llncs}

\usepackage{url}
\usepackage[colorlinks,bookmarksopen,bookmarksnumbered,citecolor=red,urlcolor=red]{hyperref}

\usepackage{amsmath}
\usepackage{cleveref}%cleveref must be loaded after amsmath
\crefname{lstlisting}{listing}{listings}
\Crefname{lstlisting}{Listing}{Listings}

\usepackage[utf8]{inputenc}

\usepackage{graphicx}
\usepackage{algorithm}
\usepackage{algorithmic}
\usepackage{mathtools}
\usepackage{colortbl}
\usepackage{xcolor}
\usepackage{hhline}

\usepackage{comment}
\usepackage{tabularx}
\usepackage{multirow}
\usepackage{bigdelim}
\usepackage{subfig} %%mai 2019
\usepackage{array}
\usepackage{tikz}
\usetikzlibrary{shapes,arrows,calc}

\usepackage{pifont}

\definecolor{Gray}{gray}{1}

\newlength{\elementssize}
\usepackage{todo}

\let\llncssubparagraph\subparagraph
\let\subparagraph\paragraph
\usepackage{titlesec}
\titlespacing*{\section}
{0pt}{2.0ex plus 1ex minus .2ex}{1.0ex plus .2ex minus .2ex}
\titlespacing*{\subsection}
{0pt}{1.3ex plus .5ex minus .2ex}{.6ex plus .2ex minus .2ex}

\let\subparagraph\llncssubparagraph

\title{{\LARGE White Paper}\\[4mm]  
Workshop on Large-scale Parallel Numerical Computing Technology (LSPANC 2020)\\[2mm]
{\large HPC and Computer Arithmetic toward Minimal-Precision Computing\\[2mm]
held at RIKEN Center for Computational Science, Japan}
}

\author{Roman Iakymchuk\inst{1,3}, Daichi Mukunoki\inst{2}, Artur Podobas\inst{2}, Fabienne J\'ez\'equel\inst{1,4}, Toshiyuki Imamura\inst{2}, Norihisa Fujita\inst{5}, Jens Huthmann\inst{2}, Shuhei Kudo\inst{2}, Yiyu Tan\inst{2}, Jens Domke\inst{2}, Kai Torben Ohlhus\inst{6}, Takeshi Fukaya\inst{7}, Takeo Hoshi\inst{8}, Yuki Murakami\inst{9}, Maho Nakata\inst{10}, Takeshi Ogita\inst{6},  Kentaro Sano\inst{2}, Taisuke Boku\inst{5}}

\institute{
Sorbonne Universit\'e, CNRS, LIP6, France %75252
\and
RIKEN Center for Computational Science, Japan
\and
Fraunhofer ITWM, Germany %67663
\and 
Universit\'e Panth\'eon-Assas, France %75231 
\and
University of Tsukuba, Japan
\and
Tokyo Woman’s Christian University, Japan
\and
Hokkaido University, Japan
\and
Tottori University, Japan
\and
University of Aizu
\and
RIKEN, Japan
}

\date{January 29 – 30, 2020}

\usepackage{graphicx}

\usepackage{cite}

\begin{document}

\maketitle

\begin{abstract}
In numerical computations, precision of floating-point computations is a key factor to determine the performance (speed and energy-efficiency) as well as the reliability (accuracy and reproducibility). However,  precision generally plays a contrary role for both. Therefore, the ultimate concept for maximizing both at the same time is the minimal-precision computing through precision-tuning, which adjusts the optimal precision for each operation and data. Several studies have been already conducted for it so far (e.g. Precimoniuos and Verrou), but the scope of those studies is limited to the precision-tuning alone. Hence, we aim to propose a broader concept of the {\it minimal-precision computing system} with precision-tuning, involving both hardware and software stack.

In 2019, we have started the Minimal-Precision Computing project to propose a more broad concept of the minimal-precision computing system with precision-tuning, involving both hardware and software stack. Specifically, our system combines (1) a precision-tuning method based on Discrete Stochastic Arithmetic (DSA), (2) arbitrary-precision arithmetic libraries, (3) fast and accurate numerical libraries, and (4) Field-Programmable Gate Array (FPGA) with High-Level Synthesis (HLS).

In this white paper, we aim to provide an overview of various technologies related to minimal- and mixed-precision, to outline the future direction of the project, as well as to discuss current challenges together with our project members and guest speakers at the LSPANC 2020 workshop; \url{https://www.r-ccs.riken.jp/labs/lpnctrt/lspanc2020jan/}.
\end{abstract}

%\begin{itemize}
%    \item go through these 3-4 main topics
%    \item then go through challenges and try to present solutions or % leave them as open questions
%\end{itemize}

\section{Introduction}
At the LSPANC 2020 workshop, we gather experts in high-performance computing, hardware, compilers, computer arithmetic, algorithms, and numerical verification. Our main purpose is to share different views from each domain on how to make hardware, compilers and tools, as well as algorithms and numerical techniques to coupe together in order to result in better Exascale computing tomorrow. This includes energy-efficient hardware resources and usage of them via tools, but also robust and reliable solvers that are tuned for new architectures. 

We begin by outlining the main topics of the workshop in~\Cref{sec:main}. Then, we proceed in~\Cref{sec:challenges} to discussing the challenges and challenging questions. In~\Cref{sec:ourproject}, we propose our strategy based on minimal-precision computing. Finally, we summarize with an outlook in~\Cref{sec:outlook}.

\section{Main Topics}
\label{sec:main}
We have considered the main five topics at the workshop and here provide a brief overview of each of them.
\subsection{FPGA technologies: architectures, compilers, and new arithmetic formats}
The traditional supercomputing facilities have been contributing to the scientific calculations which require very high performance of floating point computation. With such a background, today’s world leading supercomputers are equipped with GPU (Graphics Processing Unit) beside of ordinary CPU (Central Processing Unit). Actually, about half of the systems in TOP-10 machines in the world are the large cluster systems with tens of thousands of GPUs. However, the request for new fields of scientific computation such as deep learning is much more complicated where the traditional simple computing power cannot cover it. One of the big change in the processor architecture is the change of floating point precision, FP16 (16-bit half precision floating point) for example. Although new generation of GPUs and CPUs are supporting such a request nowadays, we need more aggressive challenge for new system architecture not only for high performance but also for high performance per energy consumption. In our Center for Computational Sciences, University of Tsukuba, we have been researching the original technologies toward next generation accelerating supercomputing. GPU is still the main player for it, but we need to consider wider variety and possibility of other kind of accelerators. One of the key technologies for processor architecture recently focused is FPGA (Field Programmable Gate Array) where the logic circuit itself can be programmed by some specific hardware description language according to the algorithm of target application. We are building a new method to combine GPU and FPGA together in a single system to compensate the weak point of GPU to be covered by the flexibility of FPGA toward complicated algorithms and problems. As the practical testbed for this challenge, our center introduced the world first cluster combing GPU and FPGA technologies for advanced scientific research. 

Furthermore, we propose a Communication Integrated Reconfigurable CompUting System (CIRCUS) to enable us to utilize high-speed interconnection of FPGAs from OpenCL HLS. CIRCUS makes a fused single pipeline combining the computation and the communication, which hides the communication latency by completely overlapping them. We used the Cygnus supercomputer operated by Center for Computational Sciences, University of Tsukuba, for the performance evaluation. Cygnus has 64 Bittware 520N FPGA boards (2 boards / node) and FPGAs are connected by an 8x8 2D-torus FPGA network. Bittware 520N Board equips an Intel Stratix10 FPGA, 32GB DDR4 external memory, and four QSFP28 external ports supporting up to 100Gbps.

As Moore's Law is slowing down, people are looking for other methods to increase the performance of calculations. Improving upon the memory bottleneck by decreasing the precision and thus decreasing input size is one option. Another option is to increase the number of operations executed on that data by exploiting the capability for parallel computation using FPGAs. Generating these computational units for FPGAs is becoming more and more convenient with HLS compilers such as IntelHLS, VivadoHLS, LegUp and Nymble. However, {\em we do not have full control when integrating arbitrary precision operations in commercial compilers}. Hence, we propose to use Nymble with arbitrary precision operations. The goal of Nymble is provide high productivity in exploring new ways by providing high compatibility with standard C codes and OpenMP support.

The inevitable end of Moore’s law motivates researchers to re-think many of the historical architectural decisions. Among these decisions we find the representation of floating-point numbers, which has remained unchanged for nearly three decades. Chasing better performance, lower power consumption or improved accuracy, researches today are actively searching for smaller and/or better representations. Today, a multitude of different representations are found in the specialized (e.g. Deep-Learning) applications as well as for general-purpose applications (e.g. {\em posits}). However, despite their claimed strengths, alternative representations remain difficult to evaluate empirically. There are software approaches and emulation libraries available, but their sluggishness only allows the smallest of inputs to be evaluated and understood. posits is a new numerical representation, introduced by professor John Gustafson in 2017 as a candidate to replace the traditional IEEE-754 representation. {\em We present our experience in designing, building and accelerating the posits numerical representation on FPGAs on a set of small use-cases}.

\subsection{Numerical verification}
Many numerical verification algorithms are actively developed using high-level programming languages. For example the Matlab/GNU Octave software VSDP (\url{https://vsdp.github.io/}) is able to compute rigorous error bounds for conic linear programs with up to 19 million variables and 27 thousand constraints using further verification algorithms from INTLAB (\url{http://www.ti3.tu-harburg.de/intlab/}). The application to large-scale problems often requires using High-Performance Computing (HPC) systems. Those systems sometimes lack of appropriate high-level language support, offer outdated versions, or hardly allow beneficial customization of the pre-installed software, like choosing specialized BLAS/LAPACK implementations. On the other hand, porting verification algorithms to another or lower-level programming language is time consuming and error prone. To overcome these issues, a recent promising approach of using lightweight Singularity (\url{https://sylabs.io/singularity/}) containers in combination with Spack (\url{https://spack.io/}) to control software dependencies is used. For the verification algorithms all necessary software customization can be prepared and tested on a desktop PC, while the final benchmark is performed on a Singularity-supporting HPC system, which is not rare in practice.

%\subsection{Precision Auto-Tuning and Control of Accuracy}
\subsection{Numerical validation and application for precision tuning}
%\notemu{I would like to change the title as above. 'numerical validation' should be appeared at the same level as numerical verification.}
In the context of high performance computing, new architectures, becoming more and more parallel, offer higher floating-point computing power. Thus, the size of the problems considered (and with it, the number of operations) increases, becoming a possible cause for increased uncertainty. As such, estimating the reliability of a result at a reasonable cost is of major importance for numerical software. We describe the principles of Discrete Stochastic Arithmetic (DSA) that enables one to estimate rounding errors by performing all arithmetic operations several times using a random rounding mode. DSA is implemented, on the one hand, in the CADNA library (\url{http://cadna.lip6.fr}) that can be used to control the accuracy of programs in half, single, double and/or quadruple precision, and, on the other hand, in the SAM library (\url{http://www-pequan.lip6.fr/~jezequel/SAM}) that estimates rounding errors in arbitrary precision programs. {\em Most numerical simulations are performed in double precision, and this can be costly in terms of computing time, memory transfer and energy consumption}. We also present the {\em PROMISE} tool (PRecision OptiMISE, \url{http://promise.lip6.fr}), based on CADNA, that {\em aims at reducing} in numerical programs the number of {\em double precision variable declarations in favor of single precision ones}, taking into account a requested accuracy of the results. Finally, in order to combine high performance and control of accuracy in a numerical simulation, we show that the cost of rounding error estimation may be avoided if particular numerical kernels are used with perturbed input data.

\subsection{Accurate numerical libraries}
Due to the non-associativity of floating-point operations and dynamic resources utilization on parallel architectures, it is challenging to obtain reproducible floating-point results for multiple executions of the same code on similar or different parallel architectures. We address the problem of reproducibility in the context of fundamental linear algebra operations – like the ones included in the BLAS (Basic Linear Algerbra Subprograms) library – and propose algorithms that yield both reproducible and accurate results. We provide implementations in the ExBLAS library available at \url{https://github.com/riakymch/exblas}. Following the hierarchical and modular structure of many linear algebra algorithms, we leverage these results and extend them to the LU factorization and Preconditioned Conjugate Gradient (PCG) method. 

In the minimal-precision computing system, we utilize fast and accurate numerical libraries, instead of MPFR, for accelerating the portions of the computation that require high accuracy. We introduce two accurate BLAS implementations developed by us, OzBLAS and BLAS-DOT2. OzBLAS is a reproducible BLAS implementation with tunable accuracy on CPUs and GPUs. It can obtain the correctly-rounded result as well as the bit-level reproducibility using the Ozaki scheme. BLAS-DOT2 is an accurate BLAS implementation on GPUs. It computes double-precision data on two-fold (quadruple) precision using the Dot2 algorithm. Both implementations are available at \url{http://www.math.twcu.ac.jp/ogita/post-k/}.

\subsection{Mixed-precision and applications}
Mixed-precision is used in verified and numerical computations for decades. One typical example is using \texttt{twosum} or \texttt{twoprod} algorithms for summation and multiplication where the operations return both the result and the error. These outputs can be stored in floating-point expansions, arrays of numbers, to represent an unveiled sum of floating-point numbers. Recently, the mixed-precision computing also started looking in the direction of reducing precision to eliminate underutilization of floating-point formats as well as to reduce energy footprint of computations. Thus, we cover both directions in this workshop.

Semidefinite programming is an important optimization problem, and higher precision than binary64 (double precision) is required for several applications. We implemented and evaluated a binary128 version of semidefinite programming solver on PC, a step toward to use hardware-implemented binary128 on FPGAs.

The GMRES(m) method is one of typical iterative methods for solving a linear system with an unsymmetric sparse coefficient matrix. Based on the restart technique employed in GMRES(m), a mixed-precision variant of GMRES(m) is easily derived. We focus on GMRES(m) using FP64 and FP32, and report the experimental evaluation of its convergence property.

In addition to rather classic mixed-precision, we also cover some benchmarks like HPC-AI as well as a study on the need of double precision in scientific applications. HPL-AI is a new benchmark program for supercomputers which is released by Jack Dongarra at ISC 2019 with its significant performance rate, 445 PFlop/s, tested on the world’s fastest supercomputer, Summit. The program measures the computation time to solve a large linear system, which is same as the well-known HPL, but it allows to use the mixed-precision techniques followed by the iterative refinements to take the advantage of the hardware capability like the 16bit floating-points which is also used in the emerging AI workloads. Unfortunately, such lower-precision computation arouses problems like the numerical instability, and even worse, causes programmers to cheat however they are not intended to do. Hence, we show examples of failures in the HPL-AI implementation to discuss with the problems for using the lower- and mixed-precision computation in scientific computations.

Among the common wisdom in High-Performance Computing is the applications’ need for large amount of double-precision support in hardware. Hardware manufacturers, the TOP500 list, and legacy software have without doubt followed and contributed to this view. In this talk, we challenge this wisdom, and we do so by exhaustively comparing a large number of HPC proxy applications on two processors: Intel’s Knights Landing (KNL) and Knights Mill (KNM). Although similar, the KNL and KNM architecturally deviate at one important point: the silicon area devoted to double-precision arithmetics. This fortunate discrepancy allows us to empirically quantify the performance impact in reducing the amount of hardware double-precision arithmetic. With the advent of a failing of Moore’s law, our results partially reinforce the view taken by modern industry (e.g., Fujitsu’s ARM64FX CPU) to integrate hybrid-precision hardware units.

\section{Challenges}
\label{sec:challenges}
\begin{comment}
TODO: Check the slides of discussions too.
\begin{enumerate}
    \item Mixed-precision computations: how to store data in order to avoid type conversion and/or duplicates?
    \item Lossy compression combined with mixed-precision
    \item How to make FPGA-programming easier and user friendly?
    \item What is your impression of stochastic arithmetic, e.g. Promise? How do you see it will apply in your fields?
    \item Will the weak reproducibility provided via minimal precision computing be suitable for your computations? 
    \item How to enable arbitrary precision support on FPGAs by compilers?
    \item For standard FP computations, user can choose among few precisions. How to help users to decide on mantissa and exponent length in case of arbitrary precision computing?
    \item Alternatively to the IEEE formats, researchers can try posits for their computations, which works well on up to IEEE single precision. Will application developers be willing to test this idea?
\end{enumerate}
\end{comment}

\subsection{Mixing precision without side effects}
Using mixed-precision approaches become very appleaing in the recent years: Langou et al.~\cite{Langou06exploitingthe} propose to solve a linear system of equations using single precision and then improve its accuracy via iterative refinement using double precision for computing residual. Haidar et al.~\cite{Haidar_ICCS_2018} extended this idea to half precision. Carson and Higham~\cite{cahi18} propose a profound theoretical study with ranges of condition numbers for solving a linear system using the LU factorization with iterative refinement and  preconditioned GMRES in three IEEE precisions, namely half, single, and double. Note the GMRES-IR uses GMRES-based iterative refinement using LU factors as preconditioners to generate a sequence of approximations. Regarding emlpoyed precisions, $u$ is the precision at which the data $A, b$ and the solution $x$ are stored; $u_f$ is the precision at which the factorization of $A$ is computed; $u_r$ is the precision at which residuals are computed.

Using multiple precisions can work nicely theoretically, however in practise code developers often need to have multiple allocations/ copies of the same data for different precisions. This leads to obvious pollution in terms of storage. Hence, our open questions are
\begin{itemize}
    \item How to avoid duplicates/ copies of data?
    \item How to eliminate type conversion?
\end{itemize}
One possible option is to store data in chunks where each number is divided into multiple chunks. For instance, double precision numbers are divided into four chunks of size 16 bits. This is one way that requires to be solidly validated in coding. However, it has a potential to eliminate duplicates and to avoid type conversion.

\subsection{FPGA programming: easier and more user friendly}
While learning to program FPGAs using some simple tests like matrix-vector multiplication, we observe the power of FPGAs but also disclose some difficulties. At first, programming FPGAs is quite different from standard CPUs or GPUs and requires some time for getting used. Second, compilation time, while using VivadoHLS, is huge (1-2 hours) even for such a small examples. Hence, we have been thinking how to facilitate FPGA programming as well as reduce by orders of magnitude its compilation time. These questions are certainly open, but are worth to be mentioned here as they intend to make the newcomers aware of possible difficulties. 

There are also a set of compilers, code generators, tools for FPGAs and it would be very helpful to get a quick cheatsheet on them or a brief guide highlighting their strengths and weaknesses. We do an attempt on this below.

\subsection{Strengths and weaknesses of SPGen, Nymble, and FloPoCo}
FloPoCo, SpGEN, and Nymble are all tools that automatically generate user-specific hardware. The main difference is \textit{where} and at \textit{what} level of abstraction the tool operates at.

At the lowest level -- and the one closest to hardware -- we find arithmetic unit generators such as FloPoCo~\cite{flopoco} or PosGen~\cite{podobas2018hardware}. These tools aims at forming the basic blocks for use in hardware, and are often used in conjunction with other high-level synthesis tools (or for inclusion in e.g. soft-cores). FloPoCo is a tool for automatically generating arithmetic units (e.g. addition, multiplications, and divisions) that use different numerical representations, and has support for both IEEE-754 and posits~\cite{gustafson2017beating}. PosGen is a similar tool but exclusively for Posit arithmetic. Both feature a rich variety of options, such as the ability to change the number of bits allocate for mantissa, exponent, or region (in the case of  Posits)-- properties that significantly impact the silicon cost of the intended arithmetic unit. FloPoCo also has support for creating simple (control-flow free) data-paths, such as, for example, creating a component that evaluates the expression: $y = x^2 + y^2 + z^2$.	

Moving up one abstraction we find data-path synthesizer tools, such as SPGen~\cite{spgen}. SPGen is built using a custom domain-specific language (DSL) which allows the user to express functions (through equations) and limited form of control-flow. The SPGen compiler can leverage arithmetic unit generators (e.g. FloPoCo) to build high-performance data-paths. Another distinction is that SPGen -- unlike e.g. FloPoCo -- generate circuitry with flow-control, which are often required for operation in real systems, and to support programming models such as stream-computing or data-flow computing. Often, SPGen relies on external components such as Direct Memory Access (DMA) modules to stream data through it. Supporting alternative numerical representation or variable precision can conceptually be done inside the SPGen DSL.

At the highest abstraction-layer we find High-Level Synthesis tools that operate on (or near) the general programming language level, such as on a sub-set of C, C++, or Java. Nymble is such a tool that builds on-top of LLVM to provide user-friendly compilation from C/C++ code. Unlike the previous two categories -- both of which depends on external units for operation -- Nymble if self-contained and can orchestrate its own execution. Nymble supports the concept of threads through a subset of the OpenMP~\cite{dagum1998openmp} 4.0 accelerator model, which simplifies porting of existing high-performance applications into it. Conceptually, Nymble can leverage both SPGen and (indirectly or directly) FloPoCo during hardware generation. Supporting and expressing alternative numerical representations or variable precision can conceptually be done by extending C/C++ construct semantics, or by introducing new variables and data-types.

\begin{figure}
    \centering
\vspace*{-3mm}     
    \includegraphics[width=1.0\textwidth]{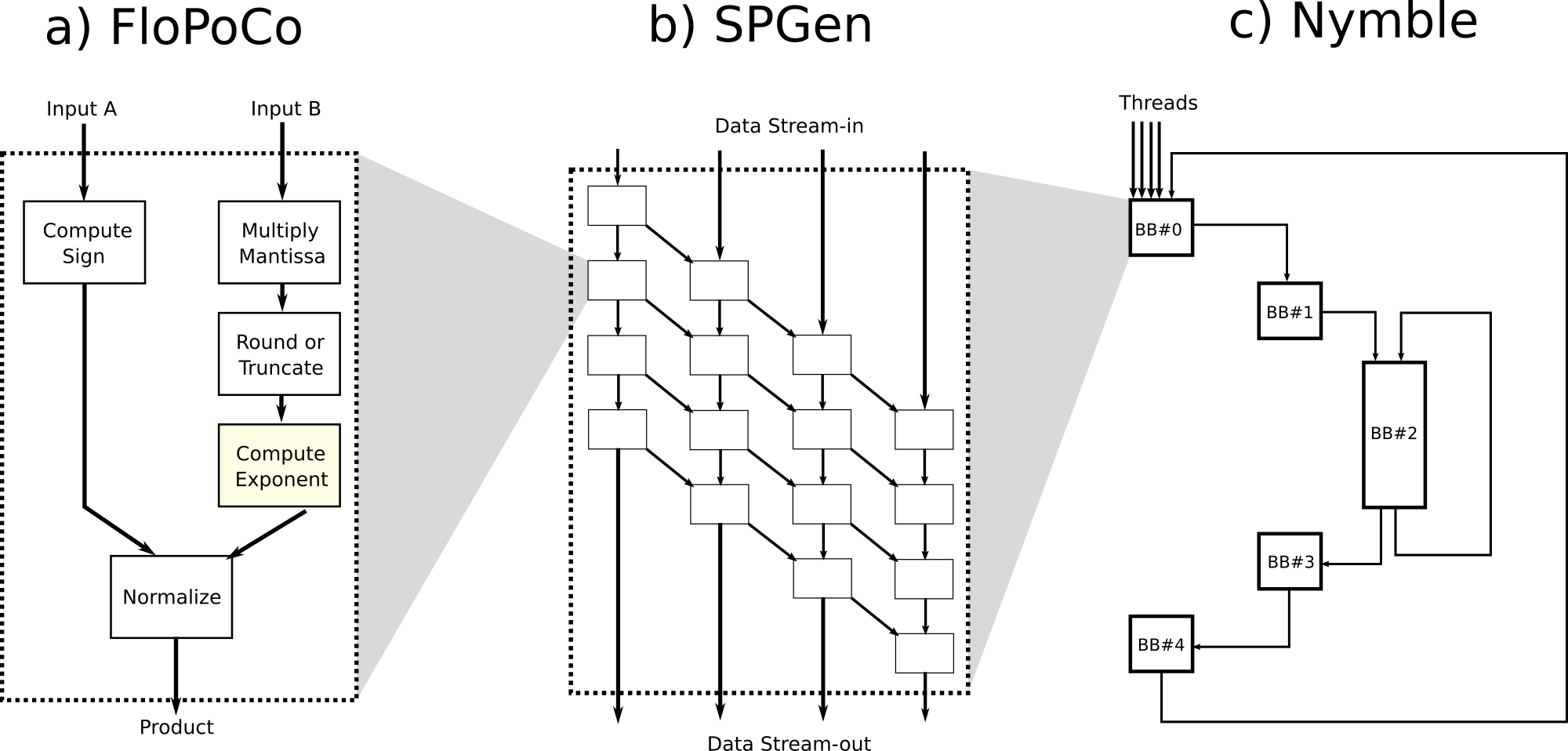}
    \caption{Conceptual picture showing the relationship between FloPoCo, SPGen, and Nymble. Here, FloPoCo (a) would be used to create the basic arithmetic hardware components for different operations. Next, (b) SPGen would use generated operations to assemble them into a data-flow pipeline. Finally, (c) Nymble would use the previously generated data-paths to create application-wide hardware. }
    \label{fig:my_label}
\vspace*{-9mm}    
\end{figure}
\begin{table}[!ht]
\vspace*{-8mm}
\caption{Strengths and weaknesses of FLoPoCo, SPGen, and Nymble; `$+$' stands for strength and `$-$' for weakness.}
\label{tb:compare3}
\begin{tabular}{l || c | c | c }
          & \textbf{FLoPoCo} & \textbf{SPGen} & \textbf{Nymble}  \\ \hline \hline
\textbf{`$+$'} & \begin{minipage} [t] {0.3\textwidth}
            \begin{itemize}
            \item High performance
            \item Flexible number representation and precision
            \end{itemize}
            \end{minipage}
          &  \begin{minipage} [t] {0.3\textwidth}
            \begin{itemize}
            \item Good performance
            \item Number representation can be (relatively) easily added
            \end{itemize}
            \end{minipage}
&  \begin{minipage} [t] {0.3\textwidth}
            \begin{itemize}
            \item Easy-to-use with C/C++
            \item Support for parallel models (e.g. OpenMP)
            \end{itemize}
            \end{minipage}            
          \\ \hline
\textbf{`$-$'}  & \begin{minipage} [t] {0.3\textwidth}
            \begin{itemize}
            \item Cannot be used stand-alone
            \item Requires hardware knowledge to integrate/use
            \end{itemize}
            \end{minipage}
          & \begin{minipage} [t] {0.3\textwidth}
            \begin{itemize}
            \item Non-standard DSL language (hard to use)
            \item Requires some hardware knowledge to use
            \end{itemize}
            \end{minipage}
          & \begin{minipage} [t] {0.3\textwidth}
            \begin{itemize}
            \item Integration of variable-precision not yet there
            \item Performance is harder to obtain (vs SPGen)
            \end{itemize}
            \end{minipage}  \\ \hline
\end{tabular}
\vspace*{-3mm} 
\end{table}

\Cref{fig:my_label} overviews the relationship between the different approaches, and we see that each increase in abstraction can conceptually re-use and leverage the benefits of the previous. For example, SPGen can use FloPoCo-generated arithmetic units, and Nymble can leverage SPGen-created data-paths.
\Cref{tb:compare3} briefly summarizes the strengths and weaknesses of each approach.

{\em To sum up:} Ideally, for design-space exploration of numerical representations and variable precision, an framework such as Nymble should be used. Here, LLVMs C/C++ front-end could be extended to support new (or arbitary) data-types, and Nymble would use FloPoCo to support these new data-types. The end-user can then describe (or re-use existing benchmark) in plain C (with support for OpenMP) and relatively easily empirically observe the effect of varying precision (or representation).

SPGen can (and should) be used when the application matches the SPGen programming model (stream-computing) and when high-performance is required; also, extending SPGen to include variable precision and alternative representations should be simpler than extending the LLVM infrastructure.

\subsection{Arbitrary precision numbers}
When standard floating-point formats may not always match our needs and help to safe the storage or energy, we may also consider arbitrary precision numbers. Such numbers allow us to literally decided on the amount of bits in every operand in every operation. However, such diversity also imposes a challenge of handling operations on such diverse data. For instance, on FPGAs such operation has to be programmed separately. Thus, we propose to agree in advance on the distance between arbitrary precision numbers (finer than floating-point numbers) in order to facilitate programming operations on FPGAs. Such stepping can be 8 or 16 bits depending on the need, bearing in mind proper and efficient memory usage. Another possibility will be to really  upon the MPFR library for arbibrary precision computations. Here, we propose to write a wrapper on top of it in order to specify mantissa and exponent sizes.

Furthermore, there is one very relevant questions regarding arbitrary precision: we are more used to standard floating-point computation with four IEEE 754 formats. When it comes to arbitrary precision, how to help users to choose the mantissa and exponent length? It is difficult to give a general answer. However, starting from the same precision as your data storage is a good idea and then trying to reduce precision on non-critical parts and increase on computationally critical ones.

\subsection{Lossy and lossless compression}
Over several years and especially now while preparing for Exascale, we have heart many times that moving data among nodes is harmful. Scientists tried to address this via communication reducing and avoiding algorithms~\cite{Carson:EECS-2015-179,CGdeeppipelines} that require communication every s iterations or fuse some iterations into one. 
The other approach is to revisit computation and usage of double precision; and try to, hence, drop precision and, thus, communicate less data. One more possibility is to compress data and then communicate using, e.g., lossy or lossless compression algorithms. This together with enhance collective communication tends to lead to smaller amount of data as well as better scaling communication.

All the above discussions and ideas, especially mixed-precision, can be enhanced with the help of compression algorithms.

\subsection{Alternatives to IEEE formats}
We are living in a quickly changing environment when both hardware and software as well as standards change and adjust. One of such changes can be observed on arithmetic formats commonly known as the IEEE 754 formats. In 2014, John Gustafson presented his idea, called unam, at the SCAN conference. Then he was invited to the ARITH 2015 conference and presented the idea there. Now Gustafson has releazed posits, unam III, as an enhanced format. Studies~\cite{florentposits19,duebenposits19} suggest that the format covers better numbers close to zero and is comparable or better than IEEE for up to single precision. Hence, posits attain good attention from DL community as well as weather forecast applications. Nearly two years ago, bfloat16 was announced that covers wider dynamic range but has smaller mantissa and thus again better suited for DL/ML codes. bfloat16 is well adapted by Intel, ARM, and Google.  Potentially, there can be more experimental or well in production arithmetic formats suited for specific needs, but not yet disclosed.

These alternatives are worth exploring although the standard computer arithmetic tools and compilers may not immediately support them. 

\section{Minimal-Precision Computing System}
\label{sec:ourproject}
\begin{figure}[ht]
\centering
\vspace*{-4mm}
\includegraphics[width=\linewidth]{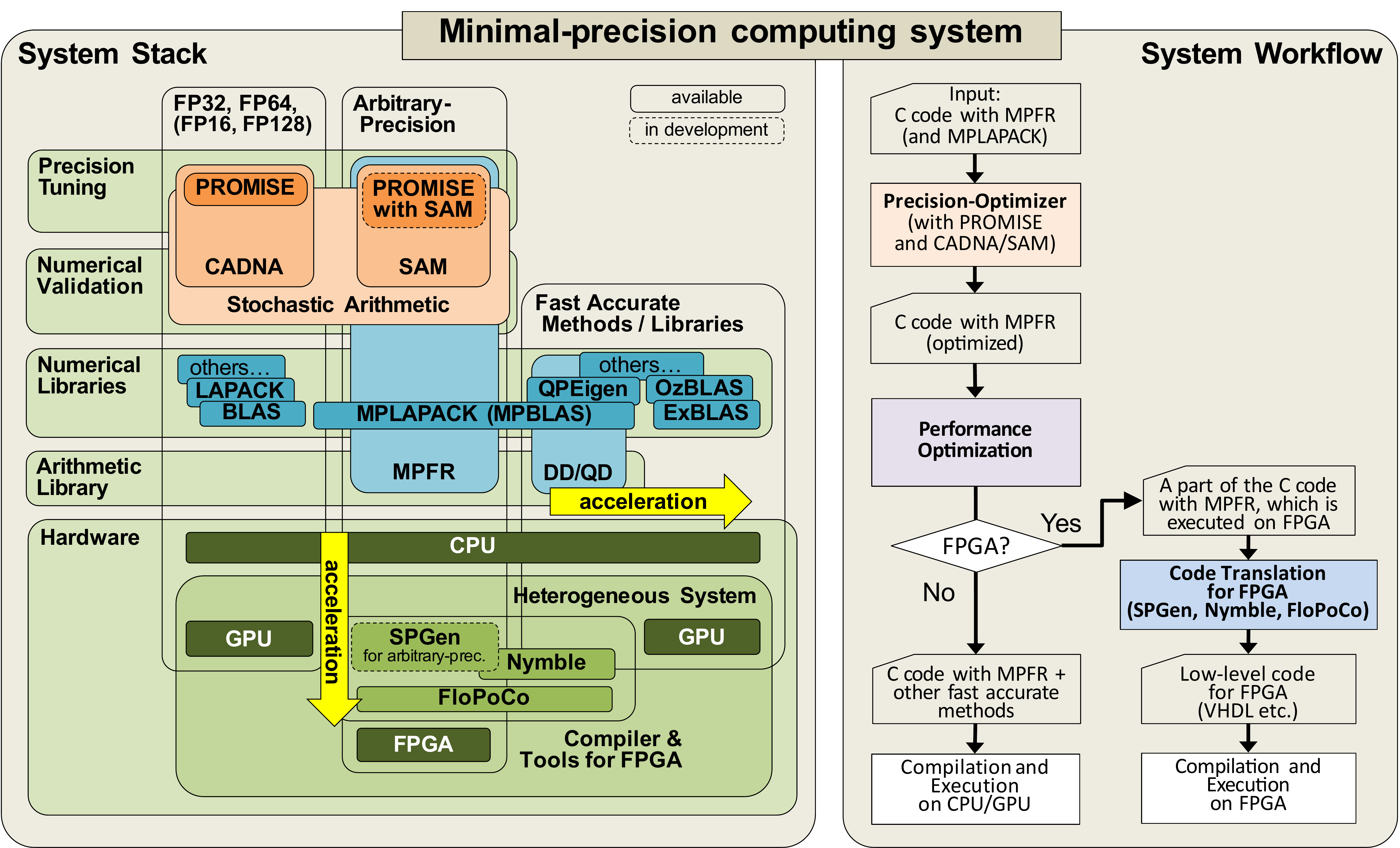}
\caption{Minimal-precision computing system overview.}
\label{fig:system}
\vspace*{-4mm}
\end{figure}

Here we would like to provide our view on the minimal-precision computing system, see~\Cref{fig:system}. The proposed system~\cite{sc19poster} combines (1) a precision-tuning method, (2) arbitrary-precision arithmetic libraries, (3) fast and accurate numerical libraries, and (4) heterogeneous architectures with Field-Program\-mable Gate Array (FPGA). 

We explain the overall procedure below.
\begin{enumerate}
\renewcommand{\labelenumi}{(\arabic{enumi})}
	\setlength{\leftskip}{-15pt} 
	\setlength{\itemsep}{0pt}      %2. ブロック間の余白
	\setlength{\parskip}{0pt}      %4. 段落間余白．
	\setlength{\itemindent}{0pt}   %5. 最初のインデント
	\setlength{\labelsep}{3pt}     %6. item と文字の間
\item We target both IEEE-754 2008 floating-point as well as arbitrary precision numbers. An input C code and a requested accuracy are given by the user. We assume that the floating-point variables and operations in the code are defined using the GNU Multiple Precision Floating-Point Reliable (MPFR) library~\cite{mpfr}. For codes using FP32/FP64, we can also rely upon MPFR or MPFR-nize them. For instance, for linear algebra operations, we can utilize MPLAPACK \cite{mplapack} -- a multi-precision Linear Algebra PACKage (LAPACK) including Basic Linear Algebra Subprograms (BLAS) based on some high-precision arithmetic libraries including MPFR.
\item The precision-tuner determines the optimal precisions for all variables in the code, which are needed to achieve the computation result with the requested accuracy. Tuning is performed by comparing with a result validated by Discrete Stochastic Arithmetic (DSA). Thus, the optimized code is reliable. Simply speaking, DSA estimates the rounding errors of floating-point operations with the guarantee of 95\% by executing the same code three times with random-rounding (randomly round-down or -up). Then, the common digits in the three results are assumed to be a reliable result. It is a general scheme applicable for any floating-point operations: no special algorithms and no code modification are needed. 
We propose to use two DSA libraries, namely CADNA~\cite{cadna} and SAM~\cite{sam}, as well as a precision tuner called PROMISE~\cite{promise}. Besides, it can be performed at a reasonable cost in terms of both performance and development  compared to the other numerical verification or validation methods. 
\item The tuned-code (with MPFR) proceeds to the performance optimization phase (and execution). At this stage, if possible to speed up some portions of the code with some fast computation methods (including GPU acceleration), those parts are replaced with them. The method must be at least as accurate as that of the required-precision. We may be able to use hardware-native floating-point operations (e.g., FP16/FP32/FP64), fast high-precision arithmetic libraries, and accurate numerical libraries. For instance, ExBLAS~\cite{exblas} and OzBLAS~\cite{ozblas} libraries for  accurate and reproducible BLAS. We assume that this step is processed manually for now, but we plan to automate or assist it. 
\item Another important possibility for performance improvement is utilizing FPGA. FPGA enables us to implement and perform arbitrary-precision floating-point operations: it realizes the ultimate mini\-mal-precision computing and achieves better performance and energy-efficiency than software implementations on general processors. Owing to the  High-Level Synthesis (HLS) technology, we can use FPGA through existing programming languages such as C/C++ and OpenCL. As a target platform, we plan to utilize the Cygnus supercomputer at the University of Tsukuba that is equipped with GPUs and FPGAs.
\end{enumerate}

\section{Outlook}
\label{sec:outlook}
At the LSPANC 2020 workshop, we aimed to gather experts in high-performance computing, hardware, compilers, computer arithmetic, algorithms, and numerical verification in order to see how each of us helps to make the Exascale computing tomorrow. HPC experts approach this issue by developing more energy-efficient hardware and corresponding software stack, including programming models to master hardware heterogeneity. Computer arithmeticians and applied mathematicians are focused on minimizing the communication overhead in numerical methods as well as optimizing the working precision but still ensuring the high quality of the final result. Hence, it was very fruitful to exchange ideas from each domain and to brainstorm on common issues using different approaches and different points of view. We have collected a set of challenges and open questions. We tried to propose solutions to each of them, however some of them still require more finer answers or strategies.

%\makeatletter
%\def\thebibliography#1{\section*{References\@mkboth
%      {REFERENCES}{REFERENCES}}\list
%      {[\arabic{enumi}]}{\settowidth\labelwidth{[#1]}\leftmargin\labelwidth
%	\advance\leftmargin\labelsep
%	\usecounter{enumi}}
%	\def\newblock{\hskip .11em plus .33em minus .07em}
%	\sloppy\clubpenalty4000\widowpenalty4000
%	\sfcode`\.=1000\relax}
%\makeatother
\bibliographystyle{plain}
%\bibliography{references}

\end{document}